\documentclass[11pt]{article}
\usepackage{moriond,epsfig}

\def\be{\begin{equation}}
\def\ee{\end{equation}}
\def\bea{\begin{eqnarray}}
\def\eea{\end{eqnarray}}

\begin{document}
\vspace*{4cm}
\title{
EVOLUTION OF GALAXY CLUSTERS IN $\Lambda$MDM COSMOLOGIES
}

\author{ 
N.A. ARHIPOVA$^{~1}$, 
T. KAHNIASHVILI$^{~2,3}$,  
V.N. LUKASH$^{~1}$
 }

\address{
$^1$Astro Space Center of P.N.Lebedev Physical Institute, 
Moscow, Russia\\
$^2$Center for Plasma Astrophysics, Abastumani
Astrophysical Observatory, Tbilisi, Georgia\\
$^3$Department of Physics and Astronomy,
Rutgers University,  Piscataway, NJ, USA\\
}

\maketitle\abstracts{
The time evolution of galaxy cluster abundance is used 
to constrain cosmological parameters in dark matter models 
containing a fraction of hot particles (massive neutrino). 
We test the modified MDM models with cosmic gravitational waves 
which are in agreement with observational data at $z=0$, and
show that they do not pass the cluster evolution test and 
therefore should be ruled out. 
The models with a non-zero cosmological 
constant are in better agreement with the evolution test.
We estimate $\Omega_\Lambda$ and find that it is strongly 
affected by a small fraction of hot dark matter:
$0.4 <\Omega_\Lambda <0.8\;$ for $\;\Omega_H /\Omega_M <0.2$.
}

\section{
Introduction
}
Any realistic cosmological model should be consistent with
observation data on large scale structure (LSS) in the Universe
in the range of scales from galaxy~\cite{Peacock} ($\sim 1 {h}^{-1}$ Mpc)
up to large scale CMB anisotropy (CMBA)~\cite{cobe} 
($\sim 1000 {h}^{-1}$ Mpc).
 
Nowadays, the most popular cosmological models are 
cold dark matter with non-zero cosmological 
constant ($\Lambda$CDM)~\cite{lcdm_1} and mixed dark matter 
without~\cite{mdm_1} and with~\cite{vkn} cosmological 
constant, where MDM is in the form of non-baryonic 
hot and cold collisionless particles. 

All cosmological models were re-addressed after CMBA experimental 
detection~\cite{cobe} to reveal their positive and negative features. 
Both $\Lambda$CDM and MDM models have met several difficulties. 

As far as $\Lambda$CDM models are concerned 
~\cite{kgb93,eke,liddle96} they demand a high value 
of the cosmological constant ($\Omega_\Lambda \geq 0.7$). 
In this case $\Lambda$CDM is able to fit a set of LSS 
observational constraints whereas at small scales 
it overproduces the number of collapsed objects 
by a factor 2 in comparison with the corresponding 
number of gravitationally bounded objects in galaxies 
cataloges~\cite{kph96}.

Regarding standard MDM models the difficulties are related with 
late galaxy(quasar) formation~\cite{pogsta} and too high number 
density of galaxy clusters at $z=0$ ~\cite{ma,shandarin}: 
standard MDM is ruled out at $2\sigma$ CL. 

One of the possibilities to overcome the sMDM difficulties 
is a consideration of MDM with some amount of cosmic
gravitational waves (CGW)~\cite{ma,alm,mikhe}. 
The importance of fundamental CGW has been discussed earlier 
in papers~\cite{staro,vsk,lm}. 
Although $\Lambda$CDM models with $\Omega_M \leq 0.3$ 
normalized by COBE 4-year data~\cite{BW} are consistent 
with the cluster number density test, in order 
to archive an agreement with CMBA and cluster abundance data in MDM models 
(for both blue ($n >1$)~{\footnote{$n$ is the slope-index of 
post-inflationary density perturbation power spectrum.}} 
and red ($n <1$) scalar perturbation spectra) 
it is necessary to take into account the CGW contribution 
in the derived value of $\Delta T/T$ at $10^0$ angular scale. 
The latter is estimated by parameter T/S, the ratio of 
the tensor to the scalar mode contributions.

As an alternative to $\Lambda$CDM and MDM, the MDM models with a 
non-zero cosmological constant ($\Lambda$MDM) have been proposed in 
\cite{knv,ndgla,prigro}. 
The advantage of these models is in retaining the inflationary 
paradigm and flat Harrison-Zel'dovich spectrum ($n=1$) with a smaller 
value (comparing with $\Lambda$CDM) of cosmological $\Lambda$-term. 
Other notable features of $\Lambda$MDM are the possibility of 
negligible CGW contribution (T/S =0) and a small fraction of 
hot particles: even $10\%$  of massive neutrinos 
($\Omega_H/\Omega_M\sim 0.1$) could change the value of cosmological 
constant, being however in good agreement with other
independent tests (CMB data~\cite{hu}, QSO lensing~\cite{kochanek}, 
Hipparcos data~\cite{hypar}, SNIa~\cite{sn}).

In this paper we consider MDM (with CGW) and $\Lambda$MDM 
(with T/S=0), applying to these models the cluster evolution test.
The latter test constraints effectively parameter $\Omega_M$ (and
hence $\Omega_\Lambda$ in flat models)~\cite{eke,bahfan,alk}. 
Our aim is to demonstrate how the presence of hot component  
could influence the $\Omega_\Lambda$ limits. 

We describe our models in Section 2 and the evolution test in Section 3, 
the results are summarized in Section 4. 

\section{Cosmological Models}

We assume that DM is given by mixture of CDM and HDM 
components in the flat background space. 
The free model parameters are: 
\begin{itemize}
\item
{$\Omega_M$, the total matter density in the
Universe ($\Omega_M = 1-\Omega_\Lambda =\Omega_b+\Omega_H+\Omega_C$,
the latter are density parameters of baryons, hot, 
and cold particles, respectively,} 
\item
{$\Omega_H/\Omega_M$, the fraction of hot DM},
\item
{$N_H$, the number of massive neutrino species},
\item
{$h\;$, the Hubble constant in units $100$ km s$^{-1}$Mpc$^{-1}$.}
\end{itemize}

The massive and massless neutrinos are described by the corresponding 
distribution functions which are evaluated from the Boltzmann-Vlasov 
collisionless equations. Cold particles (neutralino or hypotethical 
axions) are described as a pressureless fluid ($p_C=0$). 
Baryons and photons are treated as an ideal
hydrodynamic fluid satisfying the Euler equations of motion 
(we choose the fixed value for baryon density parameter, 
$\Omega_b=0.015/h^2$). 
All components interact with each other only gravitationally.

Assuming the power low post-inflationary density perturbations 
spectrum ($P_0(k) \infty k^n$), the final total density power spectrum
can be expressed as:
\begin{equation}
P(k,z)=Ak^n T^2(k,z) 
\left [ {g(\Omega_M(z)) \over{(1+z)g(\Omega_M)}} \right]^2
\end{equation}
where $A$ is the normalization constant, 
$T(k,z)$ is the total transfer function, and
$g(\Omega_M(z))$ is the suppression coefficient \cite{kofsta}. 
According to \cite{car} $g(\Omega_M(z))$  
can be approximated as
\begin{equation}
g(\Omega_M(z))=2.5\Omega_M(z)
\left(
\frac{1}{70}+\frac{209\Omega_M(z)}{140}-
\frac{\Omega_M^2(z)}{140} + \Omega_M^{\frac 47}(z) \right)^{-1}
\label{supp}
\end{equation}
where the current matter abundance $\Omega_M(z)$ can be written as follows: 
\begin{equation}
\Omega_M(z)=\Omega_M \frac{(1+z)^3} {1- \Omega_M+(1+z)^3 \Omega_M}\;,\;\;\;
\Omega_M\equiv\Omega_M(0)\;.
\end{equation}

We use the transfer function approximations for sCDM~\cite{transfer} 
and $\Lambda$MDM~\cite{Hu&Eisen} models. 

\section{Cluster Evolution Test}

The mass function for the gravitationally bounded halos of mass greater 
than $M$ formed in the flat Universe by redshift $z$ is given by~\cite{ps}
\begin{equation}
N(>M,z)=\int\limits_{M}^{\infty} n(M^{'},z) dM^{'}
\end{equation}
where $n(M,z)dM$ is the comoving number density of collapsed objects 
with masses lying in the interval ($M,M+dM$):
\begin{equation}
n(M,z) =
 \sqrt{\frac{2}{\pi}}\frac{\rho\delta_{c}}{M}
 \frac{1}{\sigma^{2}(R,z)} | \frac{d\sigma(R,z)}{dM} |
 e^{-\frac{\delta^{2}_{c}}{2\sigma^{2}(R,z)}}\;\;.
\label{N(>M)}
\end{equation}
$M=\frac{4}{3}\rho R^{3}$,
$\rho$ is the mean matter density, and $\delta_c$ is the critical
density contrast for a linear overdensity able to collaps. 
The rms amplitude of density fluctuation in the spheres of 
radius $R$ at redshift $z$ is related to the power spectrum as
\begin{equation}
\sigma^{2}(R,z)= \frac{1}{2 \pi^{2}}\int\limits_{0}^{\infty} 
P(k,z)|W(k,R)|^{2}k^{2}dk,
\end{equation}
where $W(kR)$ is a Fourier component of the top-hat window function:
$
W(x)=\frac{3}{x^3}(\sin{x}- x cos{x}).
$

For the matter dominated Universe ($\Omega_M=1$) $\delta_c=1.686$
(e.g. ~\cite{eke,liddle96}).
For the flat models with $\Lambda$-term $\delta_c$ depends weakly on 
the current matter abundance. 
The theoretical mass functions obtained with the help of the
Press-Schechter formalism (eqs.(4,5,6)) are in a good agreement 
with other methods including numerical simulations~\cite{efstat,ekena}.
Due to the exponential dependence (see eq.(5)) the  cluster number 
density is very sensitive to the value of $\sigma_8$~.

Rich galaxy clusters are strong X-rays sources characterised by 
the gas temperature (see the review papers at this Conference). 
With help of numerical simulation it was confirmed 
that $T_g \infty M^{2/3}$, the proportionality coefficient depends
slightly on cosmological models. Here, we make use the $T-M$ relation for 
isothermal gas given in ~\cite{eke}: 
\begin{eqnarray}
T_g= {7.75 \over \beta} \left ( {6.8 \over 5X+3} \right ) \left (
{M \over 10^{15} h^{-1} M_\odot} \right )^{{2 \over 3}} 
\left ( {\Omega_M \over \Omega_M(z)} \right )^{{1 \over 3}} 
\left ({\Delta_{cr} \over 178} \right )^{{1 \over 3}} (1+z)~\rm{Kev}
\label{Temmas}
\end{eqnarray}
where $\beta(\simeq 1)$ is the ratio of the galaxy kinetic energy
to the gas thermal energy, $X(\simeq 0.76)$ is the hydrogen mass fraction. 
The value $\Delta_{cr}$ is the ratio of a mean 
halo density (within the virial radius of collapsed object) to the 
critical density of the Universe at the corresponding redshift. 
For $\Omega_M \leq 1$, $\Delta_{cr}$ can be derived analytically 
and approximated as $\Delta_{cr}=178 \Omega_M^{0.45}$. 

As well as the cluster mass function $N(>M)\equiv N(>M,0)$, the 
X-cluster temperature function $N(>T)\equiv N(>T,0)$ is also
sensitive to $\sigma_R \equiv \sigma (R,0)$.    
The comparison of the observed temperature function~\cite{henry91} 
with cluster mass function~\cite{bahcen} shows an agreement between 
two approaches. Confronting with observational data both tests provide  
a powerful constraint on $\sigma$-value in different DM models 
~\cite{vkn,Bahcall98,novosdur,mikhe}. 
 
The cluster mass(temperature) functions and evolution tests in 
$\Lambda$CDM has been discussed in details 
(e.g.~\cite{ekeal,viana1,henry00}).
It has been shown that flat cosmological models are preferable in 
comparison with open ones. 
To archive a better consistency with cluster evolution observations 
the value of $\Omega_M$ should be lower than that obtained from 
the cluster number density test at $z=0$.

The dependence of the cluster mass(temperature) function 
on $z$ is model dependent due to the differences 
in growth rate of density perturbations. 
This is reflected in $\sigma$ time evolution related to 
the suppression coefficient $g(\Omega_M(z))$:  
\begin{equation}
\sigma (R,z)=\sigma_R\frac{g(\Omega_M(z))}{g(\Omega_M)}
\frac{1}{1+z}, 
\end{equation}
where $g(\Omega(z))$ is given by eq.(\ref{supp}).

We consider below the cosmological models containing massive 
neutrino. In this case $z$-dependence also appears in the 
transfer function. 

\section{Results and Discussion}
\subsection{MDM Models with Non-Zero Tensor Mode}

To test MDM models with zero cosmological constant we follow 
the normalization procedure given in~\cite{alm,mikhe}. 
All models are normalized by $\sigma_8$ 
by the best fit of observational present day cluster mass 
~\cite{bahcen,bahfan} and temperature~\cite{henry91,Don} functions.
The agreement with $\Delta T /T $ data 
leads to a non-zero derived parameter T/S. 
The required value of T/S is mainly determined by spectral index $n$, 
abundance of hot matter component $\Omega_H$, and Hubble constant $h$. 

Fig.1a presents the present day cluster temperature functions $N(>T)$ 
for different values of $\Omega_H$ in MDM models with CGW normalized by 
$\sigma_8 \simeq 0.52$. The normalization does not depend on $\Omega_H$, 
therefore all curves cross each other at some fixed point 
(corresponding to the mass $M$ in the sphere of radius $8h^{-1}$Mpc). 

A significant contribution of both CGW and massive neutrino is 
needed to fit the CMBA data. 
For $h=0.6$ and $\Omega_H=0.2$ models with flat spectra $n=1$, T/S=1.7; 
for blue spectra $n=1.1$, T/S=3.6; for red spectra $n=0.9$, 
T/S=0.6 \cite{mikhe}. An increase of $\Omega_H$ decreases the 
parameter T/S, however the problem arises with small-scale clustering 
and $Ly_\alpha$ cloud formation test ~\cite{gnedin}. 
At low $\Omega_H < 0.2$, neither possible changes of $h$ nor models 
with three species of massive neutrino can suppress high 
contribution of CGW to CMBA. 
Even neglecting the observational problem at high $\Omega_H\geq 0.2$ 
cannot help to achieve an agreement with cluster evolution test in 
MDM matter dominated models.

Figs. 1b, 2a, 2b present the cluster evolution in MDM models.  
None of the considered models can fit the data at high   
$z > 0.3$. We varied $\Omega_H$ (Fig.~1b.), $h$ (Fig.~2a.), 
and post-inflationary spectral index $n$ (Fig.~2b). 
All the models predict the number of galaxy clusters at high redshifts 
at least two orders of magnitude smaller than the observed one. 
This fact indicates that the evolution should be slower than that 
found in MDM models without a cosmological constant. 

\subsection{$\Lambda$MDM Models without CGW}

Let us consider flat $\Lambda$MDM models without CGW normalized by 
the COBE 4-year data. 

The best fit $\sigma_8$ obtained from nearby cluster observational 
abundance depends on $\Omega_M$ as: 
$\sigma_8(\mbox{cl})=0.52 \Omega_M^{-0.52 +0.13\Omega_M}$~ 
\cite{eke,liddle96}.
The coincidence of $\sigma_8(\mbox{cl})$ with $\sigma_8(\mbox{cmb})$ 
derived in COBE normalized $\Lambda$CDM model
would limit the parameters $\Omega_M$ ($\Omega_\Lambda$) by the 
corresponding values $\sim 0.3 (0.7)$ (e.g.~\cite{Bahcall98}).
Cluster evolution test bounds these parameters ever stronger
~\cite{henry00}, $\sim 0.2 (0.8)$.

As an example, we show the cluster evolution functions in 
$\Lambda$CDM model with $\Omega_\Lambda=0.7$ for the cluster masses 
$M \geq 8 \cdot 10^{14} M_\odot$ in the comoving radius $1.5h^{-1}$Mpc, 
as functions of $h$ (Fig.~3a) and $n$ (Fig.~3b).   
The cluster evolution is not practically affected 
by changes of the spectral index. Regarding parameter $h$, 
the models with $h\geq 0.65$ are preferable.

While the requested value of cosmological constant in $\Lambda$CDM 
models is quite high ($\Omega_\Lambda \geq 0.7$), 
a small amount of hot particles ($\Omega_H /\Omega_M < 0.2$)  
in $\Lambda$MDM models allows a consistency with observations 
for smaller $\Omega_\Lambda$ ~\cite{vkn,ndgla}. 

We performed the calculations of function $N(>T)$ for different parameters
$\Omega_\Lambda$ and $\Omega_H / \Omega_M$ (Figs.~4a,b). 
Increasing $\Omega_H/\Omega_M$ reduces the needed value of 
cosmological constant. Taking into account the $Ly_\alpha$ forest 
test, models with $\Omega_H/\Omega_M \simeq 0.1$ are preferable. 
Parameter $\Omega_M$ remains in the range $0.4 \leq \Omega_M \leq 0.6$  
in agreement with other results
(cf.~\cite{vkn,ndgla,prigro,novosdur}). 

Figs.~5-6 present the cluster evolution in $\Lambda$MDM models 
as functions of the parameters $\Omega_\Lambda$, 
$\Omega_H/\Omega_M$, and $h$.  
Increasing $\Omega_H/\Omega_M$ raises the needed value of 
$\Omega_\Lambda$ (in agreement with the cluster number density test). 
Changing $h$ does not significally influence the $\Omega_\Lambda$ bounds. 

Our investigation shows that cluster evolution test clearly 
indicate the existence of non-zero cosmological constant for a
set of spatially flat $\Lambda$MDM models with negligible 
abundance of CGW. The presence of a small amount of hot particles 
like massive neutrino reduces the required value of the
cosmological constant. For example, if massive  neutrinos 
consistute only $\sim 10\%$ of the total DM density, 
the models with $\Omega_\Lambda \simeq 0.5 \pm 0.1$ and
$h\simeq 0.65 \pm 0.05$ satisfy the observational data best.  

\section*{Acknowledgments}

The work of N.A.A. and V.N.L. was partially supported by RFBR (01-02-16274a)
and INTAS (97-1192). 
The work of T.K. was supported by the COBASE program of USNRC.

\clearpage
\newpage

\begin{figure}
\centerline{\psfig{figure=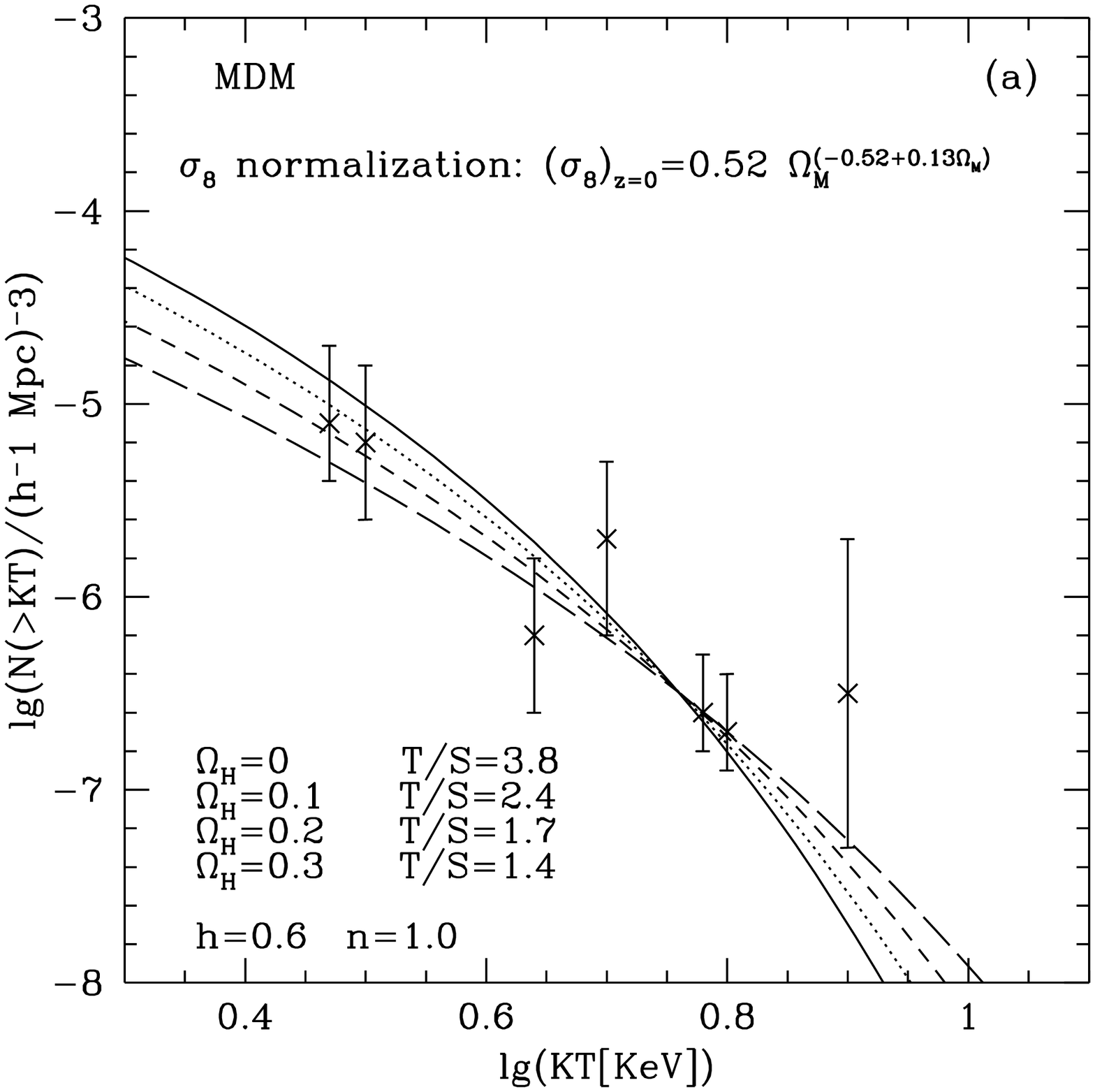,width=7cm}
            \psfig{figure=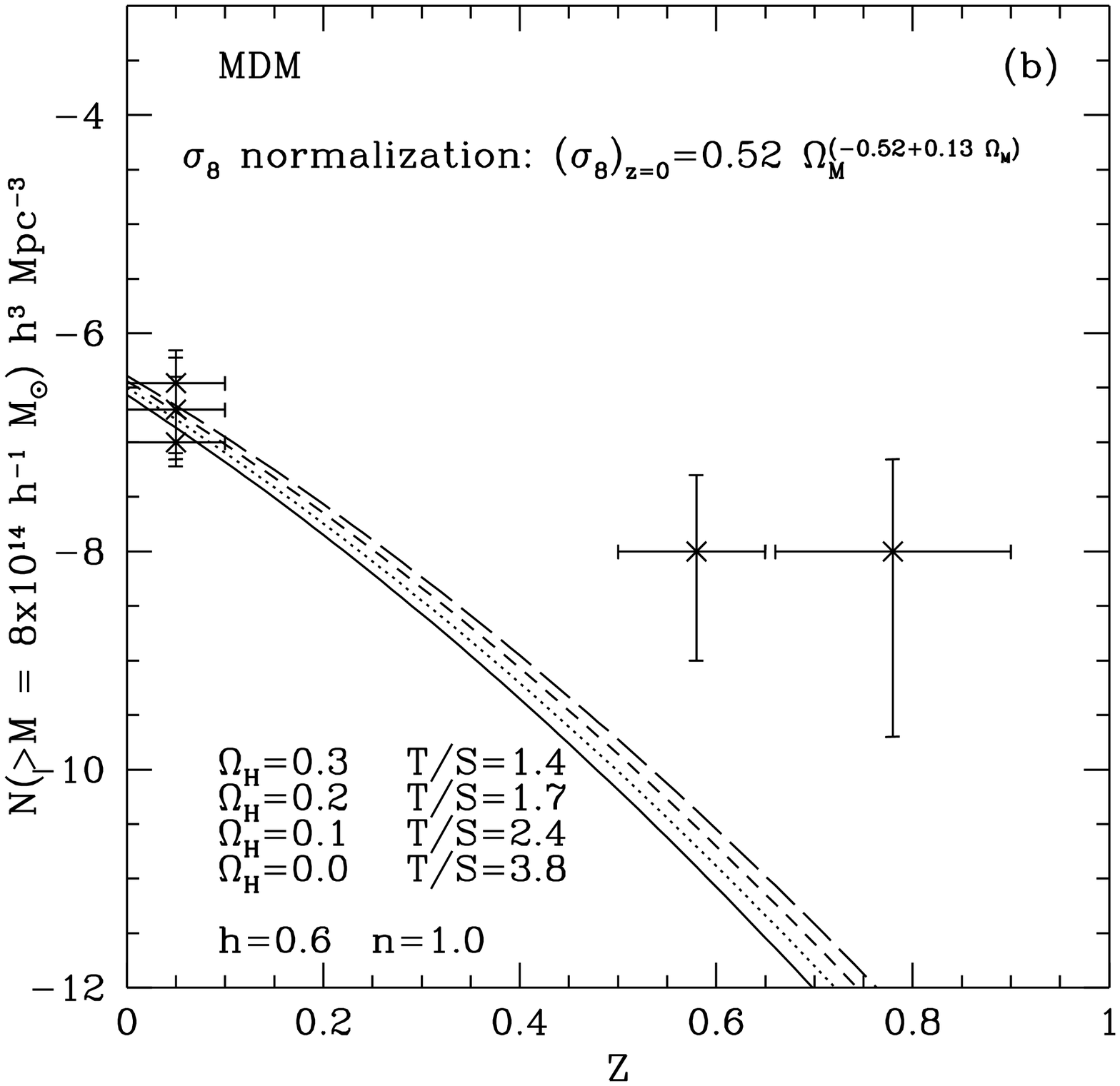,width=7cm}}
\caption{
(a) The present day cluster abundance $N(>T)$  
(data from Henry and Arnaud 1991), and (b) the cluster evolution
$N(>M = 8 \cdot 10^{14} M_\odot, z)$ (data from Bahcall and Fan 1998), 
in MDM models with CGW, normalized as $\sigma_8=0.52$, 
with $n=1.0$, $h=0.6$, $\Omega_b=0.015/h^2$, 
for $\Omega_H=0, 0.1, 0.2, 0.3$  
(solid, dot, short-dash, long dash lines, resp.).
The needed T/S  is shown.
}
\label{Fig.1}
\end{figure}

\begin{figure}
\centerline{\psfig{figure=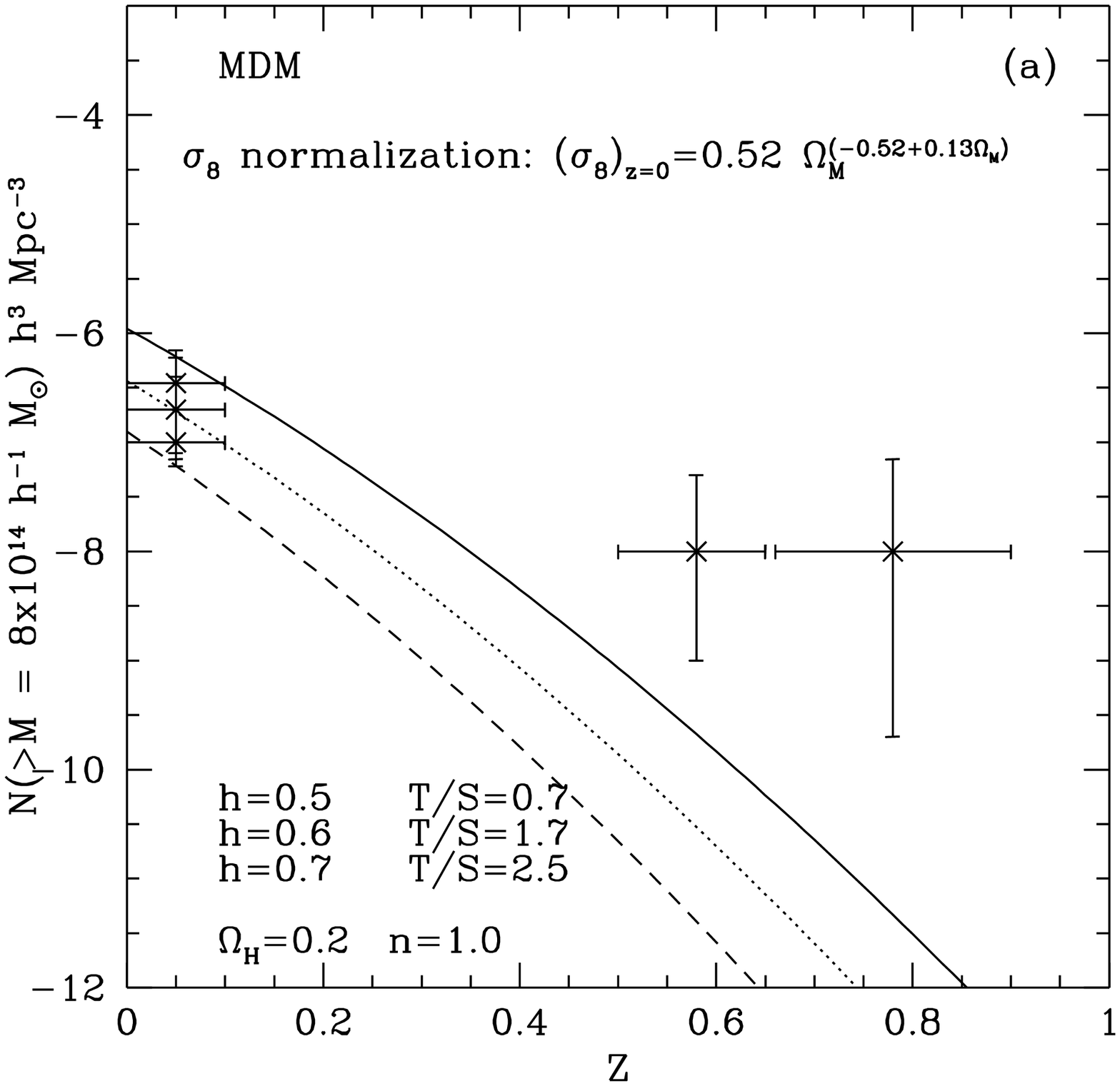,width=7cm}
            \psfig{figure=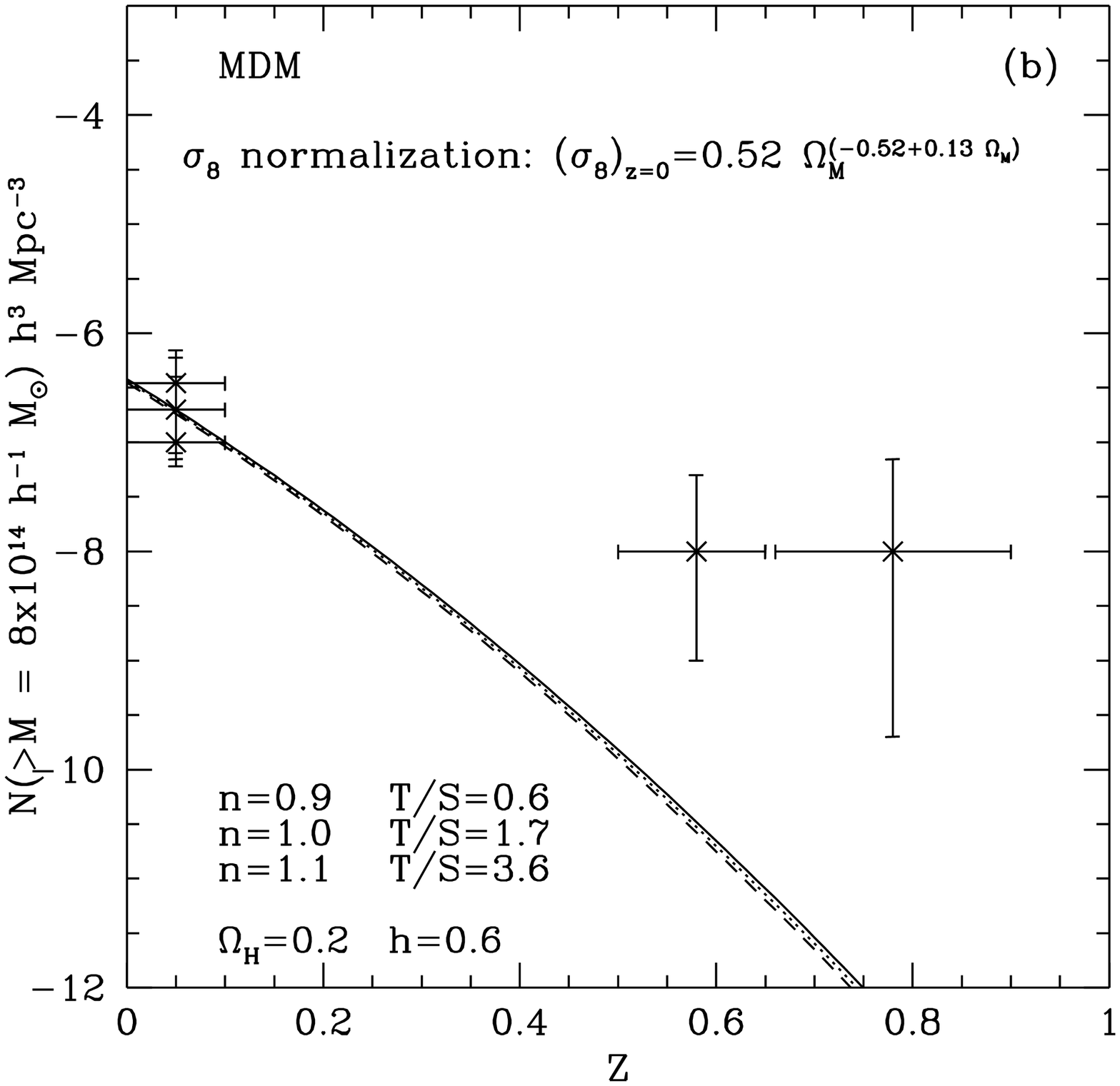, width=7cm}}
\caption{
The cluster evolution $N(>M = 8 \cdot 10^{14} M_\odot, z)$ in
MDM models normalized as $\sigma_8=0.52$ with $\Omega_H=0.2$,
$\Omega_b=0.015/h^2$, for (a) $n=1$, 
$h=0.5, 0.6, 0.7$ (solid, dot, short-dash lines, resp.), and
(b) $h=0.6$, $n=0.9, 1.0, 1.1$ (solid, dot, short-dash lines, resp.).
The needed T/S is shown. The points correspond to Bahcall and Fan 1998.
}
\label{Fig.2}
\end{figure}

\clearpage
\newpage

\begin{figure}
\centerline{\psfig{figure=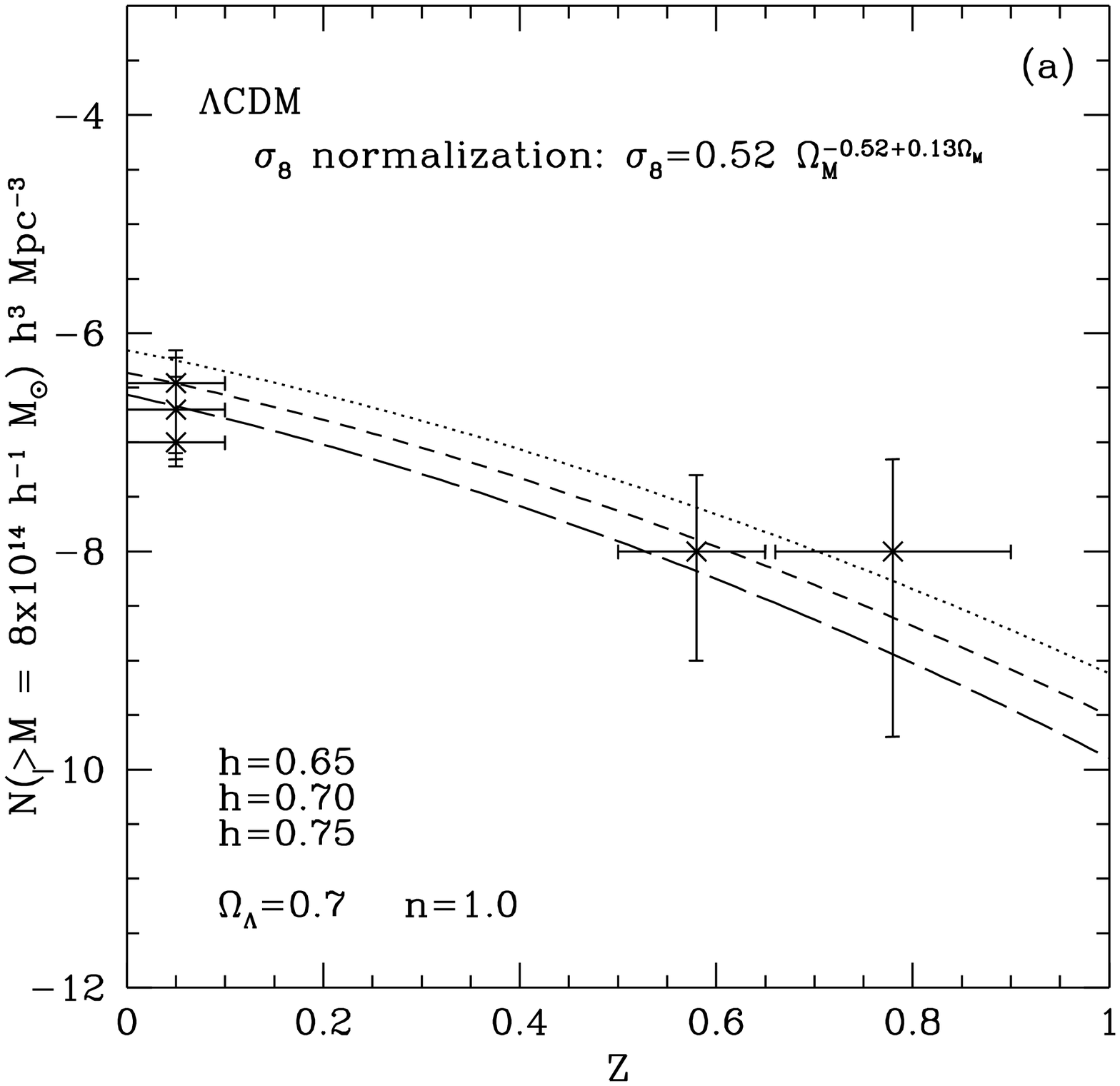,width=7cm}
\psfig{figure=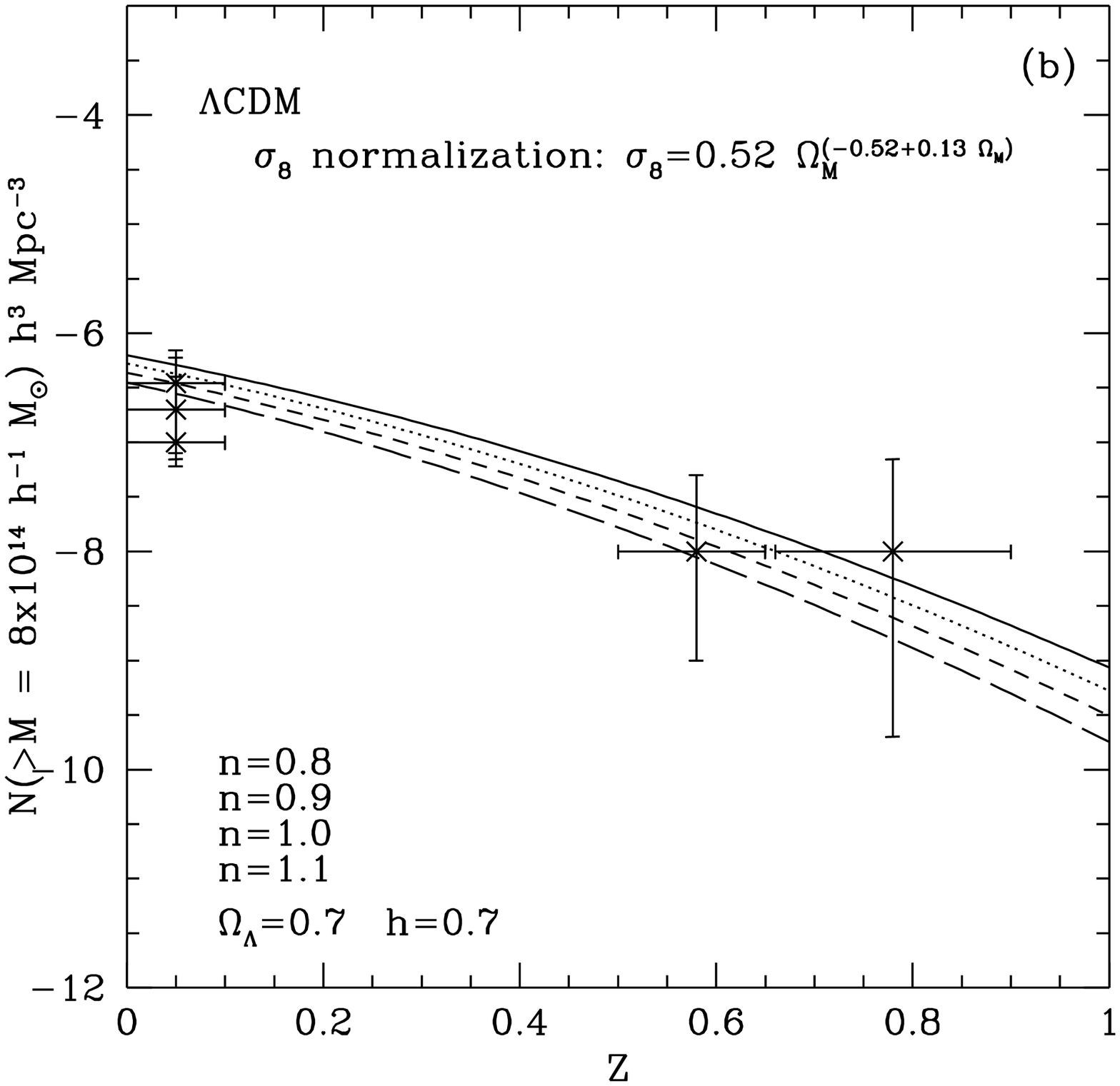,width=7cm}
}
\caption{The cluster evolution $N(>M = 8 \cdot 10^{14} M_\odot, z)$
in $\Lambda$CDM models normalized as $\sigma_8=0.52 \Omega_M^{(
-0.52\Omega_M+0.13)}$ with $\Omega_b=0.015/h^2$, $\Omega_{\Lambda}=0.7$, 
for (a) $n=1$ and $h=0.65, 0.70, 0.75$ 
(solid, dot, short-dash, long-dash lines, resp.), and 
(b) $h=0.7$, and $n=0.8, 0.9, 1.0, 1.1$ 
(solid, dot, short-dash, long-dash lines, resp.). 
The data points correspond to Bahcall and Fan 1998. } 
\label{Fig.3}
\end{figure}

\begin{figure}
\centerline{\psfig{figure=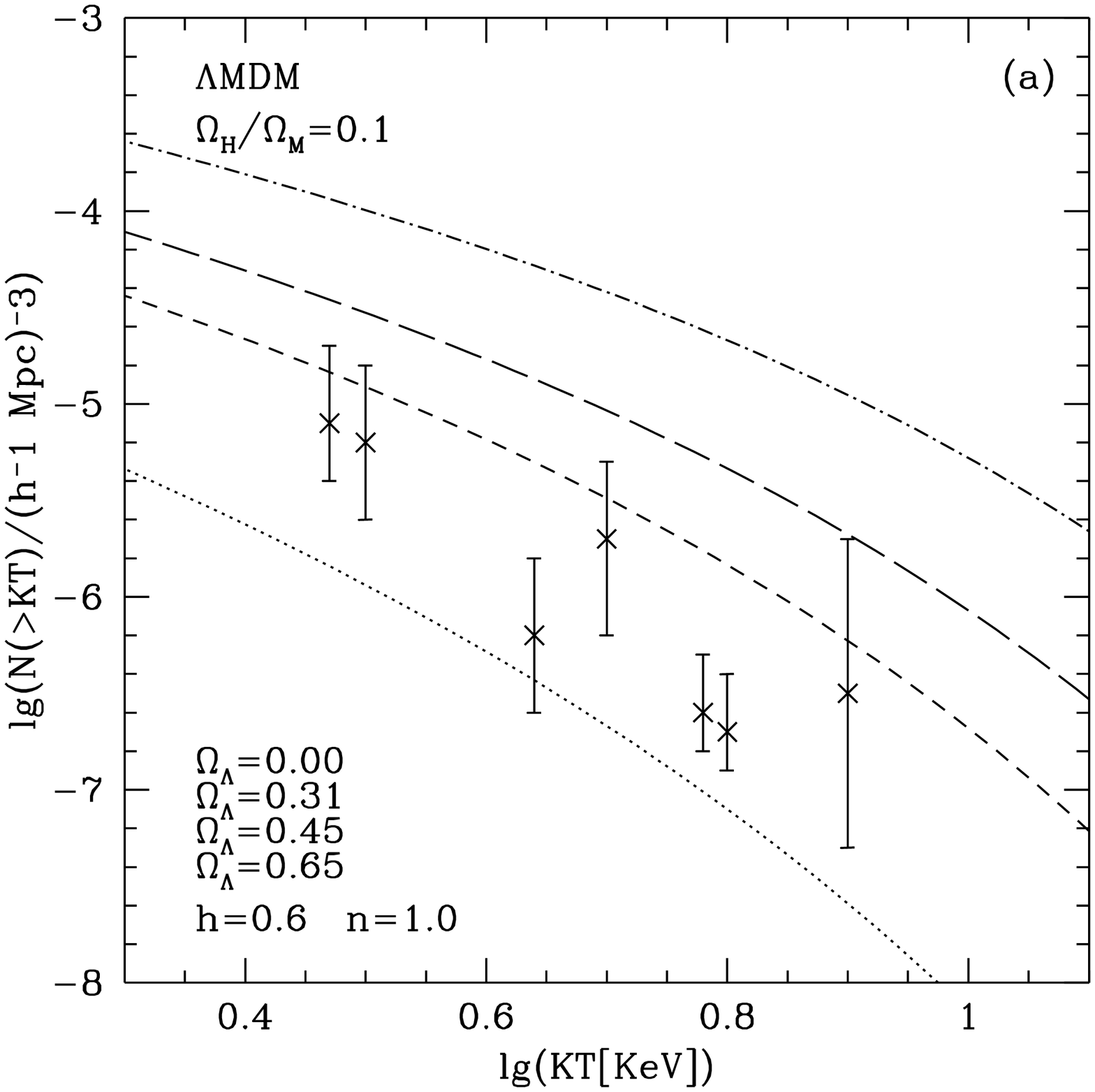,width=7cm}
\psfig{figure=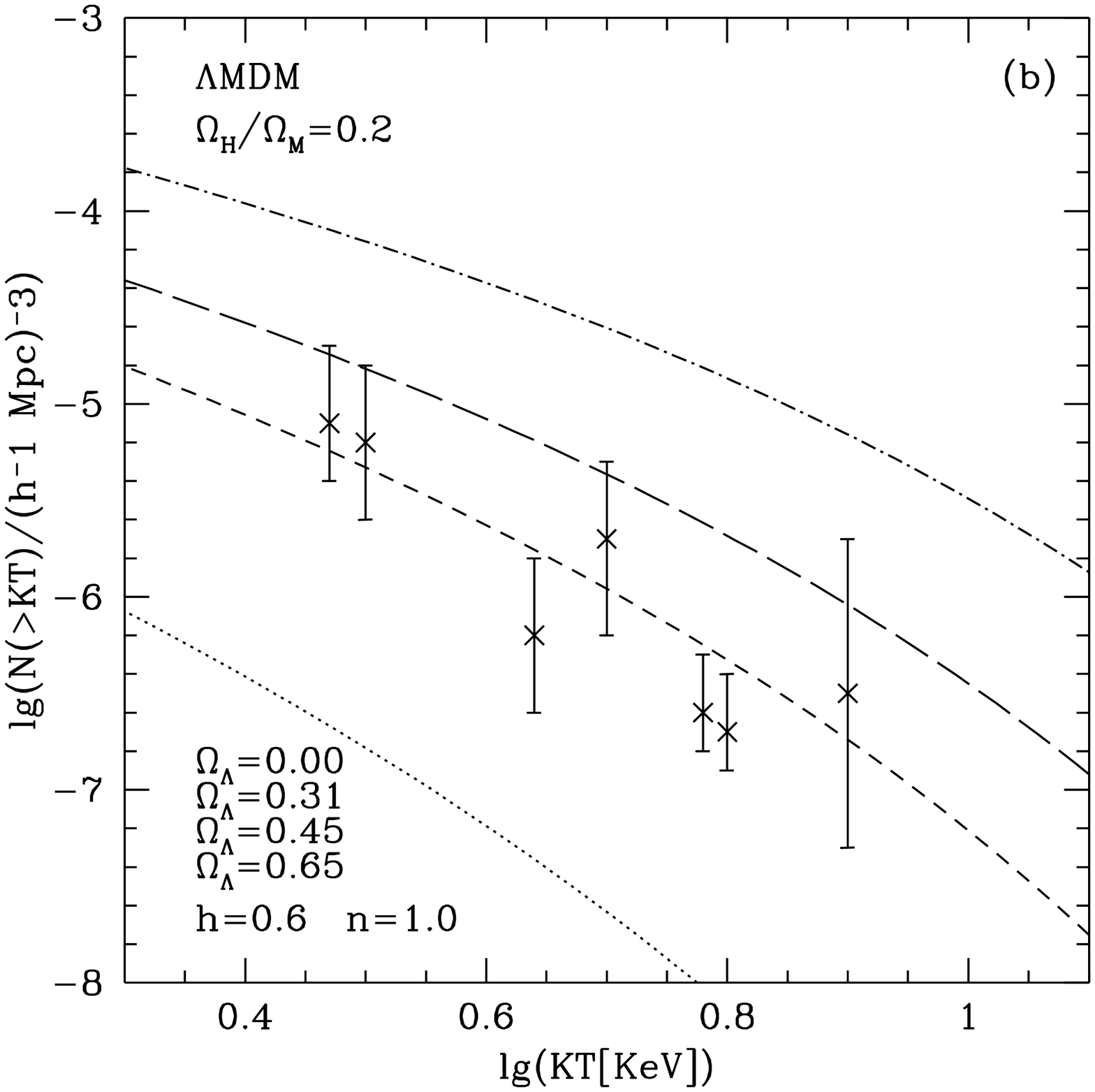,width=7cm}}
\caption{The present day cluster abundance $N(>T)$ in $\Lambda$MDM
models normalized by COBE 4-year data with $h=0.6$,
$\Omega_b=0.015/h^2$, $n=1$, for $\Omega_\Lambda=0, 0.31, 0.45, 0.65, 0.74$ 
(solid, dot, short-dash, long-dash, dot-dash lines, resp.), and
(a) $\Omega_H /\Omega_M=0.1$, 
(b) $\Omega_H /\Omega_M=0.2$. 
The points correspond to Henry and Arnaud 1991.}
\label{Fig.4}
\end{figure}

\clearpage
\newpage

\begin{figure}
\centerline{\psfig{figure=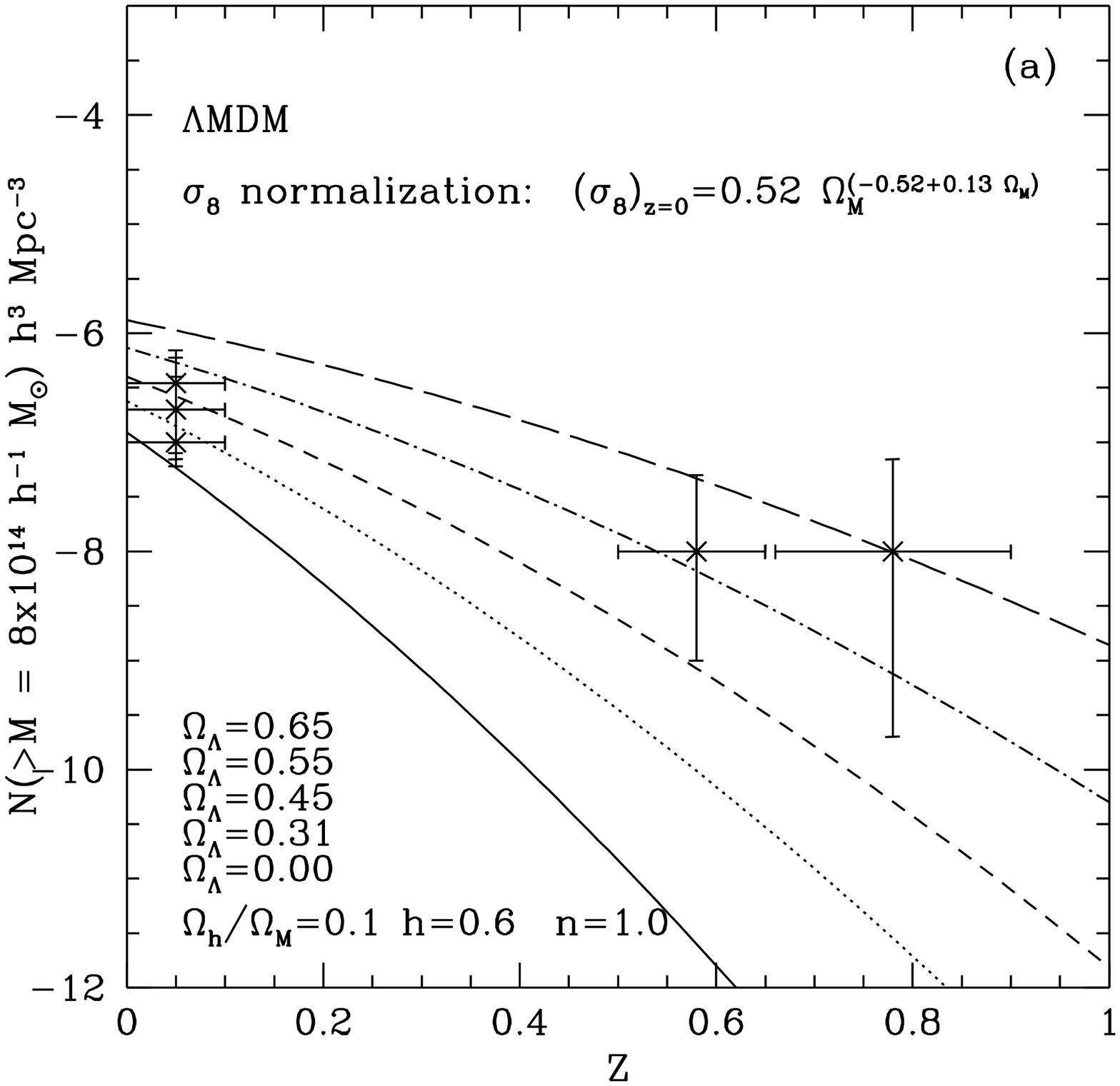,width=7cm}
          \psfig{figure=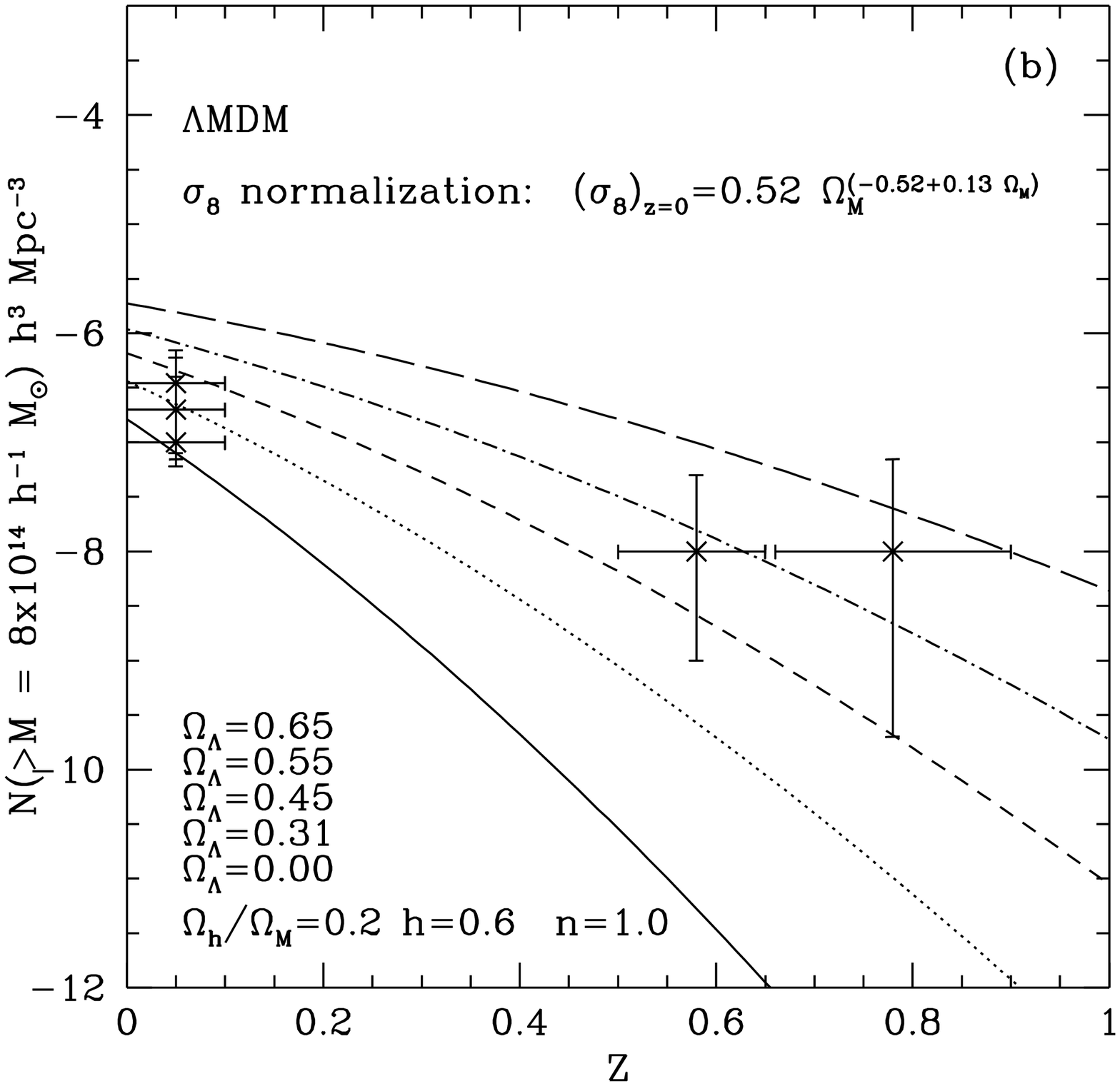,width=7cm}}
\caption{The cluster evolution $N(>M = 8 \cdot 10^{14} M_\odot, z)$ 
in $\Lambda$MDM models normalized as $\sigma_8=0.52 \Omega_M^{(
-0.52\Omega_M+0.13)}$ with  $\Omega_b=0.015/h^2$, $h=0.6$,
$n=1$,  for $\Omega_\Lambda =0, 0.31, 0.45, 0.55, 0.65$ 
(solid, dot, short-dash,  dot-dash lines, long-dash, resp.), and  
(a) $\Omega_H / \Omega_M =0.1$, 
(b) $\Omega_H / \Omega_M =0.2$. 
The data points correspond to Bahcall and Fan 1998.}
\label{Fig.5}
\end{figure}

\begin{figure}
\centerline{\psfig{figure=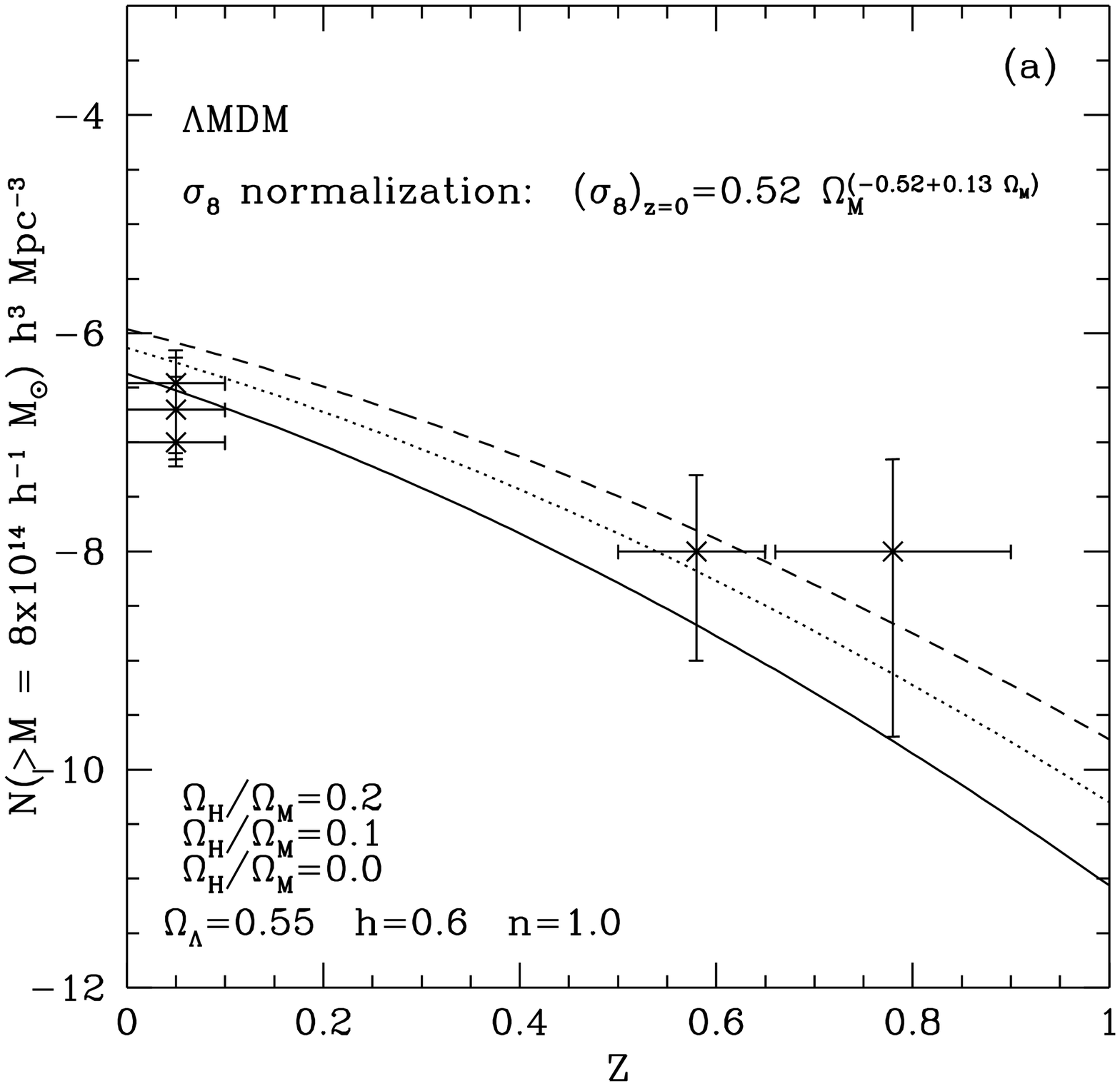,width=7cm}
          \psfig{figure=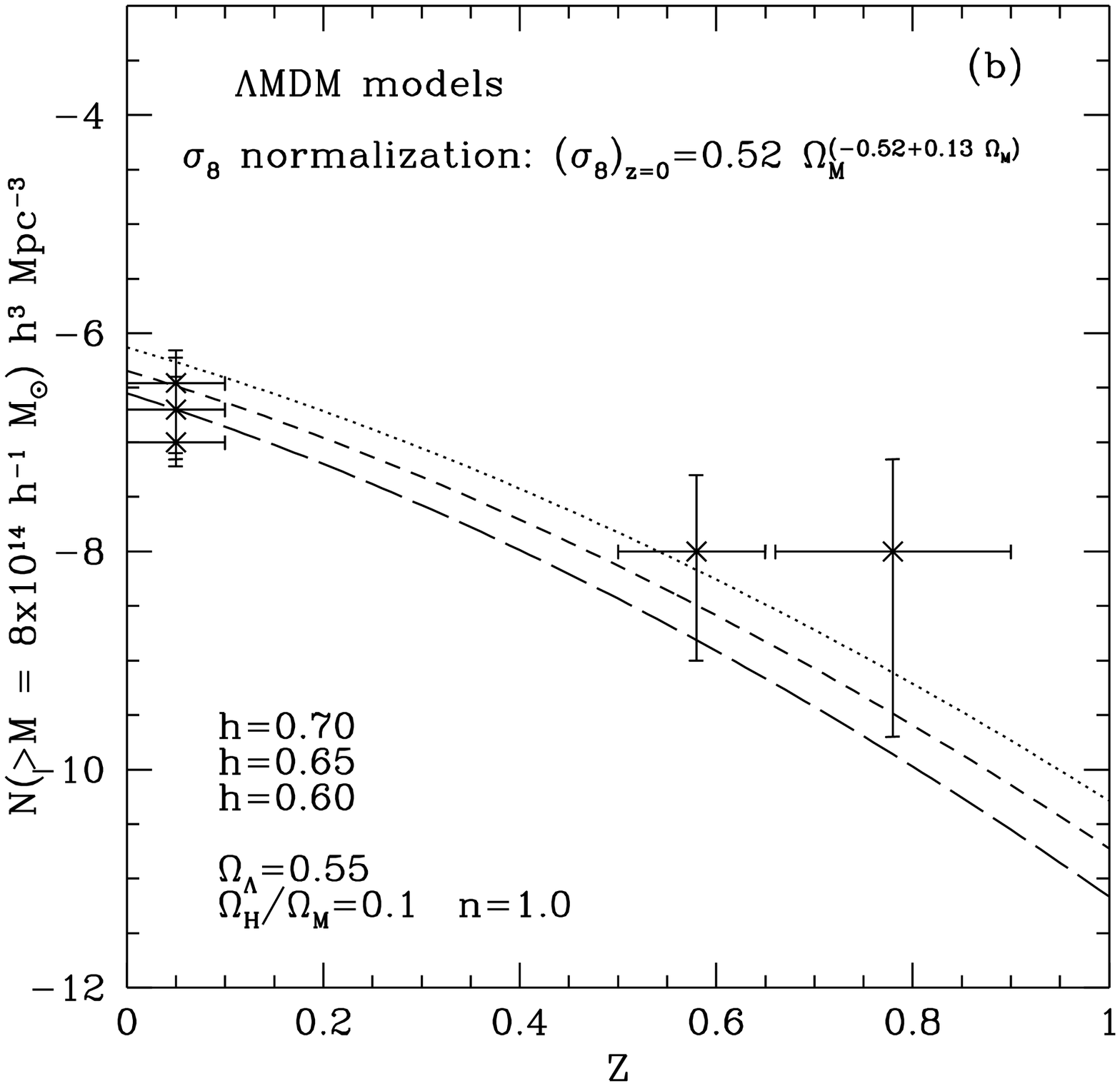,width=7cm}}
\caption{The cluster evolution $N(>M = 8 \cdot 10^{14} M_\odot, z)$
in $\Lambda$MDM models normalized as $\sigma_8=0.52 \Omega_M^{-0.52+
0.12 \Omega}$ with $\Omega_\Lambda=0.55$, $\Omega_b=0.015/h^2$, $n=1$, 
for (a) $h=0.6$, $\Omega_H / \Omega_M =0, 0.1, 0.2$ 
(solid, dot, short-dash, resp), and 
(b) $\Omega_H / \Omega_M =0.1$, $h=0.6, 0.65, 0.7$ 
(dot, short dash and long dash lines, resp.).  
 The data points correspond to Bahcall and Fan 1998.}
\label{Fig.6}
\end{figure}

\end{document}